# Quantum Chemistry for Solvated Molecules on Graphical Processing Units (GPUs) using Polarizable Continuum Models

Fang Liu,[1,2] Nathan Luehr,[1,2] Heather J. Kulik,[1,3] Todd J. Martínez[1,2]

[1]*Department of Chemistry and The PULSE Institute, Stanford University, Stanford, CA 94305*

[2]*SLAC National Accelerator Laboratory, Menlo Park, CA 94025*

[3]*Department of Chemical Engineering, Massachusetts Institute of Technology, Cambridge, MA 02139*

**Abstract:** The conductor-like polarization model (C-PCM) with switching/Gaussian smooth discretization is a widely used implicit solvation model in chemical simulations. However, its application in quantum mechanical calculations of large-scale biomolecular systems can be limited by computational expense of both the gas phase electronic structure and the solvation interaction. We have previously used graphical processing units (GPUs) to accelerate the first of these steps. Here, we extend the use of GPUs to accelerate electronic structure calculations including C-PCM solvation. Implementation on the GPU leads to significant acceleration of the generation of the required integrals for C-PCM. We further propose two strategies to improve the solution of the required linear equations: a dynamic convergence threshold and a randomized block-Jacobi preconditioner. These strategies are not specific to GPUs and are expected to be beneficial for both CPU and GPU implementations. We benchmark the performance of the new implementation using over 20 small proteins in solvent environment. Using a single GPU, our method evaluates the C-PCM related integrals and their derivatives more than 10X faster than a conventional CPU-based implementation. Our improvements to the linear solver provide a further 3X acceleration. The overall calculations including C-PCM solvation require typically 20-40% more effort than their gas phase counterparts for moderate basis set and molecule surface discretization level. The relative cost of the C-PCM solvation correction decreases as the basis sets and/or cavity radii increase. Therefore description of solvation with this model should be routine. We also discuss applications to the study of the conformational landscape of an amyloid fibril.

# 1. Introduction

Modeling the influence of solvent in quantum chemical calculations is of great importance to understanding solvation effects on electronic properties, nuclear distributions, spectroscopic properties, acidity/basicity, and mechanisms of enzymatic and chemical reactions.[1-4] Explicit inclusion of solvent molecules in quantum chemical calculations is computationally expensive and requires extensive configurational sampling to determine equilibrium properties. Implicit models based on a dielectric continuum approximation are much more efficient, and are an attractive conceptual framework to describe solvent effects within a quantum mechanical (QM) approach.[1]

Among these implicit models, the apparent surface charge (ASC) methods are popular because they are easily implemented within QM algorithms and can provide excellent descriptions of the solvation of small- and medium-sized molecules when combined with empirical corrections for non-electrostatic solvation effects.[4] ASC methods are based on the fact that the reaction potential generated by the presence of the solute charge distribution may be described in terms of an apparent charge distribution spread over the solute cavity surface. Methods such as the polarizable continuum models[5] (PCM) and its variants such as conductor-like models (COSMO,[6] C-PCM,[7] also known as GCOSMO,[8] and IEF-PCM[9-11]) are the most popular and accurate of these ASC algorithms.

While PCM calculations are much more efficient than their explicit solvent counterparts, their application in quantum mechanical calculations of large-scale biomolecular systems can be limited by CPU computational bottlenecks.[4] Graphical processing units (GPUs), which are characterized as stream processors,[12] are especially suitable for parallel computing involving massive data and numerous groups have explored their use for electronic structure theory.[13-24] Implementation of gas phase ab initio molecular calculations[19-21] on GPUs led to greatly enhanced performance for large systems.[25-26] Here, we harness the advances[27] of stream processors to accelerate the computation of implicit solvent effects, effectively reducing the cost of PCM calculations. These improvements will enable simulations of large biomolecular systems in realistic environments.

# 2. Conductor-like Polarizable Continuum Model

The original Conductor-like screening model (COSMO) was introduced by Klamt and Schuurmann.[6] In this approach, the molecule is embedded in a dielectric continuum with permittivity $\varepsilon$, and the solute forms a cavity within the dielectric with unit permittivity. In this electrostatic model, the continuum is polarized by the solute, and the solute responds to the electric field of the polarized continuum. The electric field of the polarized continuum can be described by a set of surface polarization charges on the cavity surface. Then the electrostatic component of the solvation free energy can be represented by the interaction between the polarization charges and solute, in addition to the self-energy of the surface charges. For numerical convenience, the polarization charge is often described by a discretization in terms of $M$ finite charges residing on the cavity surface. The locations of the surface charges are fixed, and the values of the charges can be determined via a set of linear equations

$$\mathbf{Aq} = -f(\mathbf{Bz} + \mathbf{c}) \qquad (1)$$

where $\mathbf{q} \in \mathbb{R}^M$ is the discretized surface charge distribution, $\mathbf{A} \in \mathbb{R}^{M \times M}$ is the Coulomb interaction between unit polarization charges on two cavity surface segments, $\mathbf{B} \in \mathbb{R}^{M \times N}$ is the interaction between nuclei and a unit polarization charge on a surface segment, $\mathbf{z} \in \mathbb{R}^N$ is the vector of nuclear charges for the $N$ atoms in the solute molecule, and $\mathbf{c} \in \mathbb{R}^M$ is the interaction between the unit polarization charge on one surface segment and the total solute density. The parameter $f=(\varepsilon-1)/(\varepsilon+k)$ is a correction factor for a polarizable continuum with finite dielectric constant. In the original COSMO paper, $k$ was set to 0.5. Later work by Truong and Stefanovic[8] (GCOSMO) and Cossi and Barone[7] (C-PCM) suggested that $k=0$ was more appropriate on the basis of an analogy with Gauss' Law. We use $k=0$ throughout in this work, although both cases are implemented in our code.

The precise form of the $\mathbf{A}$, $\mathbf{B}$, and $\mathbf{c}$ matrices/vectors depends on the specific techniques used in cavity discretization. In order to obtain continuous analytic gradients of solvation energy, York and Karplus[28] proposed the Switching-Gaussian formalism (SWIG), where the cavity surface van der Waal spheres are discretized by Lebedev quadrature points. Polarization charges are represented as spherical Gaussians centered at each quadrature point (and not as simple point charges). Lange and Herbert[29] proposed another form of switching function referred to here as "improved Switching-Gaussian" (ISWIG). Both SWIG and ISWIG formulations use the following definitions for the fundamental quantities $\mathbf{A}$, $\mathbf{B}$, and $\mathbf{c}$:

$$A_{kl} = \frac{\mathrm{erf}(\zeta'_{kl} |\vec{r}_k - \vec{r}_l|)}{|\vec{r}_k - \vec{r}_l|} \qquad (2)$$

$$A_{kk} = \frac{\zeta_k}{\sqrt{2\pi}} \mathbf{S}_k^{-1} \qquad (3)$$

$$B_{Jk} = \frac{\mathrm{erf}(\zeta_k |\vec{r}_k - \vec{R}_J|)}{|\vec{r}_k - \vec{R}_J|} \qquad (4)$$

$$c_k = \sum_{\mu\nu} P_{\mu\nu} L^k_{\mu\nu} \qquad (5)$$

$$\begin{aligned} L^k_{\mu\nu} &= -(\mu | \hat{J}^{screened}_k | \nu) \\ &= -\int \phi_\mu(\vec{r}) \frac{\mathrm{erf}(\zeta_k |\vec{r} - \vec{r}_k|)}{|\vec{r} - \vec{r}_k|} \phi_\nu(\vec{r}) dr \end{aligned} \qquad (6)$$

where $\vec{r}_k$ is the location of the $k$th Lebedev point and $\vec{R}_J$ is the location of the $J$th nucleus with atomic radius $R_J$. The Gaussian exponent for the $k$th point charge is given as

$$\zeta_k = \frac{\zeta}{R_I \sqrt{w_k}} \qquad (7)$$

where $\zeta$ is an optimized exponent for the specific Lebedev quadrature level being used (as tabulated[28] by York and Karplus) and $w_k$ is the Lebedev quadrature weight for the $k$th point. The combined exponent is then given as:

$$\zeta'_{kl} = \frac{\zeta_k \zeta_l}{\sqrt{\zeta_k^2 + \zeta_l^2}} \tag{8}$$

The atom-centered Gaussian basis functions used to describe the solute electronic wavefunction are denoted as $\phi_\mu$ and $\phi_\nu$ and $P_{\mu\nu}$ is the corresponding density matrix element. Finally, the switching function which smoothes the boundary of the van der Waals spheres corresponding to each atom (and thus makes the solvation energy continuous) is given by $\mathbf{S}_k$. For ISWIG, this switching function is expressed as:

$$\mathbf{S}_k = \prod_{J, k \notin J}^{atoms} S_{wf}(\vec{r}_k, \vec{R}_J)$$
$$S_{wf}(\vec{r}_k, \vec{R}_J) = 1 - \frac{1}{2}\left\{\mathrm{erf}\left[\zeta_k\left(R_J - |\vec{r}_k - \vec{R}_J|\right)\right] + \mathrm{erf}\left[\zeta_k\left(R_J + |\vec{r}_k - \vec{R}_J|\right)\right]\right\} \tag{9}$$

Similar, but more involved, definitions are used in SWIG (which we have also implemented, but only ISWIG will be used in this paper).

Once $\mathbf{q}$ is obtained by solving Eq. 1, the contribution of solvation effects to the Fock matrix is given by:

$$\Delta \mathbf{F}^S = \sum_{k=1}^{M} q_k L_{\mu\nu}^k \tag{10}$$

where the Fock matrix of the solvated system is $\mathbf{F}^{solvated} = \mathbf{F}^0 + \Delta \mathbf{F}^S$ and $\mathbf{F}^0$ is the usual gas phase Fock operator. This modified Fock matrix is then used for the self-consistent field (SCF) calculation.

As usual, the atom-centered basis functions are contractions over a set of primitive atom-centered Gaussian functions:

$$\phi_\mu(r) = \sum_{i=1}^{l_\mu} c_{\mu i} \chi_i(r) \tag{11}$$

Thus, the one electron integrals from Eq. 6 that are needed for the calculation of $\mathbf{c}$ and $\Delta \mathbf{F}^S$ are:

$$(\mu | \hat{J}_k^{screened} | \nu) = \sum_{i=1}^{l_\mu} \sum_{j=1}^{l_\nu} c_{\mu i} c_{\nu j} [\chi_i | \hat{J}_k^{screened} | \chi_j] \tag{12}$$

where we use brackets to denote one-electron integrals over primitive basis functions and parentheses to denote such integrals for contracted basis functions. In the following, we use the indices $\mu, \nu$ for contracted basis functions and the indices $i, j, k, l$ are used to refer to primitive Gaussian basis functions.

Smooth analytical gradients are available for COSMO SWIG/ISWIG calculations due to the use of a switching function which makes surface discretization points smoothly enter/exit the cavity definition. The total electrostatic solvation energy of COSMO is

$$\Delta G_{els} = (\mathbf{Bz})^\dagger \mathbf{q} + \mathbf{c}^\dagger \mathbf{q} + \frac{1}{2f}\mathbf{q}^\dagger \mathbf{Aq} \tag{13}$$

Thus the PCM contribution to the solvated SCF energy gradient with respect to the nuclear coordinates $R_I$ of the $I$th atom is given by

$$\nabla^*_{R_I}(\Delta G_{els}) = \mathbf{z}^\dagger(\nabla_{R_I}\mathbf{B}^\dagger)\mathbf{q} + (\nabla^*_{R_I}\mathbf{c}^\dagger)\mathbf{q} + \frac{1}{2f}\mathbf{q}^\dagger(\nabla_{R_I}\mathbf{A})\mathbf{q} \tag{14}$$

where $\nabla^*_{R_I}$ denotes that the derivative with respect to the density matrix is not included. The contribution of changes in the density matrix to the gradient is readily obtained from the gradient subroutine in vacuo (see supporting information for details).

In the COSMO-SCF process described above, there are three computationally intensive steps:

1. building $\mathbf{c}$ and $\Delta \mathbf{F}^S$ from Eqs. 5 and 10

2. solving the linear system in Eq. 1

3. evaluating the PCM gradients from Eq. 14

We discuss our acceleration strategies for each of these steps in Section 4 below.

## 3. Computational Methods

We have implemented a GPU-accelerated COSMO formulation in a development version of the TeraChem package. All COSMO calculations use parameters stated as follows unless otherwise specified. The environment dielectric constant corresponds to aqueous solvation ($\varepsilon$=78.39). The cavity uses an ISWIG[29] discretization density of 110 points/atom and cavity radii which are 20% larger than the Bondi radii.[30-32] An ISWIG screening threshold of $10^{-8}$ is used, meaning that molecular surface (MS) points with a switching function value less than this threshold are ignored. The conjugate gradient[33] (CG) method is used to solve the PCM linear equations, with our newly proposed random Jacobi preconditioner (RBJ) with block size 100. The electrostatic potential matrix $\mathbf{A}$ is explicitly stored and used to calculate the necessary matrix-vector products during CG iterations.

In order to verify correctness and also to assess performance, we compare our code with the CPU based commercial package, Q-Chem[34] 4.2. For all the comparison test cases, Q-Chem uses exactly the same PCM settings as TeraChem, except for the CG preconditioner. Q-Chem uses the "diagonal decomposition" together with a block Jacobi preconditioner based on an octree spatial partition. We use OpenMP paralellization in Q-Chem because we found this to be faster than its MPI[35] parallelized version based on our tests on these systems. In order to use OpenMP parallelization in this version of Q-Chem, we use the fast multipole method[36-37] (FMM) and the "no matrix" mode, which rebuilds the $\mathbf{A}$ matrix "on the fly."

We use a test set of six molecules (Figure 1) to investigate the relationship of the threshold and resulting error in the CG linear solve. For each molecule, we used five different structures: one optimized structure, and four distorted structures obtained by performing classical mo-lecular dynamics (MD) simulations on the first structure with Amber ff03 force fields[38] at 500K. A summary of the name, size, and preparation method for these molecules, together with coordinate files, is provided in Supporting Information (SI).

In the performance section, we select a test set of 20 experimental protein structures identified by Kulik, et al.[39] where inclusion of a solvent environment was essential to find optimized structures in good agreement with experimental results. The molecules are listed in SI and range in size from around 100 to 500 atoms. Most were obtained from aqueous solution NMR experiments. For these test molecules, we conduct a number of restricted Hartree Fock (RHF) single point energy and nuclear gradient evaluations with the 6-31G basis set.[40] These calculations are carried out in both PCM environment and in the gas phase. For some of these test molecules we also use basis sets of different sizes, including STO-3G,[41] 3-21G,[42] 6-31G*, 6-31G**,[43] 6-31++G,[44] and 6-31+G*. We use these test molecules to identify optimum algorithm parameters and to study the performance of our approach as a function of basis set size.

In the application section, we investigate how COSMO solvation influences the conformational landscape of a model protein by expansive geometry optimization with both RHF and the range corrected exchange-correlation functional ωPBEh.[45] Both of these approximations include the full strength long-range exact exchange interactions that are vital to avoid self-interaction and delocalization errors. Such errors can lead to unrealistically small HOMO–LUMO gaps.[46] We obtain seven different types of stationary point structures for the protein in gas phase and in COSMO aqueous solution ($\varepsilon=78.39$), with a number of different basis sets (STO-3G, 3-21G, 6-31G). Grimme's D3 dispersion correction[47] is applied to some minimal basis set calculations, here referred to as RHF-D and ωPBEh-D.

## 4. Acceleration strategies

*4a. Integral calculation on GPUs.*

Building **c** and $\Delta \mathbf{F}^S$ requires calculation of one electron integrals and involves a significant amount of data parallelism, making these well suited for calculation on GPUs. The flowchart in Figure 2 summarizes our COSMO-SCF implementation. Following our gas phase SCF implementation,[19,48] the COSMO related integrals needed for **c** and $\Delta \mathbf{F}^S$ are calculated in a direct SCF manner using GPUs. Here each GPU thread calculates integrals corresponding to one fixed primitive pair. However, the rest of the calculation, most significantly the solution of Eq. (1), is handled on the CPU.

From Eq. (5) and (10), it follows that the calculations for **c** and $\Delta \mathbf{F}^S$ are very similar, so one might be tempted to evaluate $L^k_{\mu\nu}$ once and use it in both calculations. In practice this approach is not efficient. Because $\Delta \mathbf{F}^S$ depends on the surface charge distribution ($q_k$) and therefore on **c** through Eq. (1), **c** and $\Delta \mathbf{F}^S$ cannot be computed simultaneously. As the storage requirements for $L^k_{\mu\nu}$ are excessive, it is ultimately more efficient to calculate the integrals for **c** and $\Delta \mathbf{F}^S$ separately from scratch.

The algorithm for evaluating $\Delta \mathbf{F}^S$ is shown schematically in Figure 3 for a system with three *s* shells and a GPU block size of 1×16 threads. The first and the third *s* shells contain 3 primitive Gaussian functions each; the second *s* shell has 2 primitive Gaussian functions. A block of size 1×16 is used for illustrative purposes. In practice, a 1×128 block is used for optimal occupancy and memory coalescence. Primitive pairs, $\chi_i\chi_j$, that make negligible contributions are not calculated and these are determined by using a Schwartz-like bound[49] with a cutoff, $\varepsilon^{screen}$, of $10^{-12}$ atomic units:

$$[ij\,|_{\text{Schwartz}} = [\chi_i\chi_j\,|\,\chi_i\chi_j]^{1/2} < \varepsilon^{screen} \quad (15)$$

The surviving pair quantities are preloaded to the GPU global memory and each is fetched by a unique GPU thread at the beginning of the integral kernel. Quantities related to each molecular surface (MS) grid point (charge $q_k$, coordinates $r_k$, switching-Gaussian exponent $\zeta_k$) are also preloaded in global memory. Each thread loops over all MS grid points to accumulate the Coulomb interaction between its primitive pair and all grid points as follows.

$$\Delta F_{ij}^S = -\sum_k q_k c_{\mu i} c_{\nu j} [\chi_i\,|\,\hat{J}_k^{screened}\,|\,\chi_j] \quad (16)$$

The result is stored to an output array in global memory. The last step is to form the solvation correction to the Fock matrix

$$\Delta F_{\mu\nu}^S = \sum_{\chi_i\chi_j \in \mu\nu} \Delta F_{ij}^S \quad (17)$$

on the CPU by adding each entry of the output array to its corresponding Fock matrix entry.

The algorithm for evaluating **c** is shown schematically in Figure 4. Although the same set of primitive integrals are evaluated as for the evaluation of $\Delta\mathbf{F}^S$, there are several significant differences. First, the surface charge density, $q_k$, is replaced by the density matrix element corresponding to each contracted pair. The screening formula can then be augmented with the density as follows.

$$[ij\,|_{\text{Schwartz}} = |P_{\mu\nu}|[\chi_i\chi_j\,|\,\chi_i\chi_j]^{1/2} \quad (18)$$

The density matrix elements are loaded with the other pair quantities at the beginning of the kernel. Second, the reduction is now carried out over primitive pairs rather than MS points. For the $\Delta\mathbf{F}^S$ kernel, the sum over MS points was trivially achieved by accumulating the integral results evaluated within each thread. For **c**, however, the sum over pair quantities would include terms from many threads, assuming pair quantities are again distributed to separate threads as in the $\Delta\mathbf{F}^S$ kernel. In this case, each thread in the CUDA block must evaluate a single integral between its own primitive pair and a common $k$th grid point. The result can then be stored to shared memory and a block reduction for the $b^{th}$ block produces the following partial sum:

$$c_k^b = -\sum_{\chi_i,\chi_j \in \text{block}(b)} P_{\mu\nu} c_{\mu i} c_{\nu j} [\chi_i\,|\,\hat{J}_k^{screened}\,|\,\chi_j]\,. \quad (19)$$

This sum is then stored in an output array in global memory of size $M \times n_b$, where $n_b$ is the number of GPU thread blocks in use and $M$ is the number of MS grid points. After looping over all MS grid points, the output array is copied to CPU, where we sum across different blocks and obtain the final $c_k = \sum_{b=1}^{n_b} c_k^b$.

Alternatively, the frequent block reductions can be eliminated from the kernel's inner loop. Instead of mapping each primitive pair to a thread, each MS point is distributed to a separate thread. Each thread loops over primitive pairs to accumulate the Coulomb interaction between its MS point and all primitive pairs, so that each entry of **c** is trivially accumulated within a single thread. This algorithm can be seen as a transpose of the $\Delta \mathbf{F}^S$ kernel and is referred to here as the "pair-driven kernel." The reduction heavy algorithm is referred as the "MS-driven kernel." Depending on the specifics of the hardware, one or the other of these might be optimal. We found little difference on the GPUs we used, and the results presented here use the MS-driven kernel.

All algorithms discussed above can be easily generalized to situations with angular momenta higher than *s* functions. In each loop, each thread calculates the Coulomb interaction between a MS point and a batch of primitive pairs instead of a single primitive pair. For instance, for an *sp* integral, each GPU thread calculates integrals of 3 primitive pairs $[\chi_s, \chi_p^x], [\chi_s, \chi_p^y], [\chi_s, \chi_p^z]$ in each loop. We wrote six separate GPU kernels for the following momentum classes: *ss*, *sp*, *sd*, *pp*, *pd*, *dd*. These kernels are launched sequentially.

*4b. Conjugate Gradient Linear Solver*

The typical dimension of **A** in Eq. (1) is $10^3 \times 10^3$ or larger. Since Eq. (1) only needs to be solved for a few right-hand sides, iterative methods can be applied and are much preferred over direct methods based on matrix inversion. Because the Coulomb operator is positive definite, conjugate gradient (CG) methods are a good choice. At the *k*-th step of CG, we search for an approximate solution $x_k$ in the *k*-th Krylov subspace $\mathcal{K}_k(A,b)$, and the distance between $x_k$ and the exact solution can be estimated by the residual vector:

$$\mathbf{r}_k = \mathbf{A}\mathbf{x}_k - \mathbf{b} \qquad (20)$$

The CG process terminates when the norm of the residual vector, ||$\mathbf{r}_k$||, falls below a threshold δ. A wise choice of δ can reduce the number of CG steps while maintaining accuracy.

The CG process converges more rapidly if **A** has small condition number, i.e. looks more like the identity. Preconditioning transforms one linear system to another that has the same solution, but is easier to solve. One approach is to find a preconditioner matrix, **C**, that approximates $\mathbf{A}^{-1}$. Then, the problem $\mathbf{C}\mathbf{A}x = \mathbf{C}b$ has the same solution as the original system but the matrix **CA** is better conditioned. The matrix **A** of Eq. (1) is often ill-conditioned because some of the diagonal elements, which represent the self-energy of surface segments partially buried in the "switching" area, are ~7-8 orders larger in magnitude than other diagonal elements.

In the following paragraphs we discuss our strategies to choose the CG convergence threshold δ and to generate a preconditioner for the linear equation Eq. (1).

*4b-i) Dynamic convergence threshold for CG*

We must solve Eq. (1) in each SCF step. The traditional strategy (referred to here as the fixed threshold scheme) is to choose a CG residual threshold value (e.g., $\delta \approx 10^{-6}$) and use this threshold for all SCF iterations. With this strategy, CG may require hundreds of iterations to converge in the first few SCF iterations for the computation of medium-sized systems (~500 atoms), making the linear solve cost as much time as one Fock build. However, in the early SCF iterations, the solute electronic structure is still far from the final solution, so it is pointless to get an accurate solvent reaction field consistent with the inaccurate electronic structure. In other words, we can use larger $\delta$ for Eq. (1) in the early stages of the SCF, allowing us to reduce the number of CG iterations (and thus the total cost of the linear solves over the entire SCF process).

The simplest approach to leverage this observation uses a loose threshold $\delta_1$ for the early iterations of the SCF and switches to a tight threshold $\delta_2$ when close to SCF convergence. The maximum element of the DIIS error matrix $\mathbf{X}^T(\mathbf{SPF-FPS})\mathbf{X}$, henceforth the "DIIS error," was used as an indicator for SCF convergence, where $\mathbf{S}$ is the AO overlap matrix[50] and $\mathbf{X}$ is the canonical orthogonalization matrix. When the DIIS error reached $10^{-3}$, we switched from the loose threshold $\delta_1$ to the tight threshold $\delta_2$ in the CG solver. We define the loose and tight thresholds according to the relation $\delta_1 = s \cdot \delta_2$, where s >1 is a scaling factor. We call this adaptive strategy the "2-$\delta$ switching threshold." Numerical experimentation on a variety of molecules showed that for reasonable values of $\delta_2$ ($10^{-5}$-$10^{-7}$), $s=10^4$ was a good choice which minimized the total number of CG steps required for an SCF calculation. The effect of the 2-$\delta$ switching threshold strategy is shown in Figure 5. The number of CG steps in the first few SCF iterations is significantly reduced, and the total number of CG steps over the entire SCF procedure is halved. However, there is an abrupt increase of CG steps at the switching point, making that particular SCF iteration expensive. In order to remove this artifact and potentially increase the efficiency, we investigated an alternative dynamic threshold strategy.

Luehr *et al*[51] first proposed a dynamic threshold for the precision (32-bit single vs. 64-bit double) employed in evaluating two-electron integrals on GPUs. We extend this idea to the estimation of the appropriate CG convergence threshold for a given SCF energy error. We use a set of test molecules (shown in Figure 1) at both equilibrium and distorted nonequilibrium geometries (using RHF with different basis sets and $\epsilon$=78.39) to empirically determine the relationship between the CG residual norm and the error it induces in the COSMO energy. We focus on the first COSMO iteration (i.e. the first formation of the solvated Fock matrix). The CG equations are first solved with a very accurate threshold for the CG residual norm, $\delta=10^{-10}$ atomic units. Then the CG equations are solved with progressively less accurate values of $\delta$ and the resulting error in the COSMO energy (compared to the calculation with $\delta=10^{-10}$) is tabulated. The average error for the six tested molecules is plotted as a function of the CG threshold in Figure 6. We found the resulting error to be insensitive to the basis set used. Therefore we used the 6-31G results to generate an empirical equation relating the error and $\delta$ by a power-law fit. We further shifted this equation above twice the standard deviation to provide a bound for the error. This fit is plotted in Figure 6 and given by:

$$Err(\delta) = 0.01 \times \delta^{1.07} \qquad (21)$$

where *Err(δ)* is the COSMO energy error. We use Eq. (21) to dynamically adjust the CG threshold for the current SCF iteration by picking the value of $\delta$ that is predicted to result in a DIIS error safely below ($10^{-3}$ times smaller than) the DIIS error of the previous SCF step. This error threshold ensures that error in CG convergence does not dominate the total SCF error. For the first SCF iteration, where there is no previous DIIS error as reference, we choose a loose threshold, δ=1. As shown in Figure 5, the number of CG steps required for each SCF iteration is now rather uniform. This strategy efficiently reduces CG steps without influencing the accuracy of the result. As shown in Figure 7, this approach typically provides a speed-up of 2X to 3X for systems with 100-500 atoms.

### *4b-ii) Randomized block-Jacobi preconditioner for CG*

York and Karplus[28] proposed a symmetric factorization, which is equivalent to Jacobi preconditioning. Lange and Herbert[52] later used a block Jacobi preconditioner, which accelerated the calculation by about 20% for a large molecule. Their partitioning scheme (referred to as octree in our later discussion) of the matrix blocks is based on the spatial partition of MS points in the fast multipole method (FMM),[36-37] implemented with an octree data structure. Here we propose a new randomized algorithm, which we refer to as RBJ, to efficiently generate the block diagonal preconditioner without detailed knowledge of the spatial distribution of surface charges. The primary advantage of the RBJ approach is that it is very simple to generate the preconditioner, although it may also have other benefits associated with randomized algorithms.[53] As we will show, the performance of the RBJ preconditioner is at least as good as the more complicated octree preconditioner.

Since $\mathbf{A} \in \mathbb{R}^{m \times m}$ is symmetric, there exists some permutation matrix $\mathbf{P}$ such that the permuted matrix $\mathbf{PAP}$ is block-diagonal dominant. The block-diagonal matrix, $\mathbf{M}$, is then constructed from $l \times l$ diagonal blocks of $\mathbf{PAP}$, and can be easily inverted to obtain $\mathbf{C} = \mathbf{PM}^{-1}\mathbf{P} \approx \mathbf{A}^{-1}$ as a preconditioner of $\mathbf{A}$. We generate the permutation matrix $\mathbf{P}$ in the following way: at the beginning of the CG solver, we randomly select a pivot $A_{kk}$, sort the elements of the *k*th row by descending magnitude, pick the first $l$ column indices and form the first diagonal block of $\mathbf{M}$ with the corresponding elements, repeating the procedure for the remaining indices until all rows of $\mathbf{A}$ have been accounted for. The inverse $\mathbf{M}^{-1}$ is then calculated and its non-zero entries (diagonal blocks) are stored and used throughout the Block Jacobi preconditioned CG algorithm.[54]

The efficiency of the RBJ preconditioner depends on the block size. As block size increases, more information about the original matrix $\mathbf{A}$ is kept in $\mathbf{M}$, and the preconditioner $\mathbf{C}$ becomes a better approximation to $\mathbf{A}^{-1}$. Thus, larger block sizes will lead to faster convergence of the CG procedure, at the cost of expending more effort to build $\mathbf{C}$. In the limit where the block size is equal to the dimension of $\mathbf{A}$, $\mathbf{C}$ is an exact inverse of $\mathbf{A}$ and CG will converge in 1 step. However, in this case, building $\mathbf{C}$ is as computationally intensive as inverting $\mathbf{A}$. We find that a block size of 100 is usually large enough to get significant reduction in the number of CG steps required for molecules with 100-500 atoms at a moderate discretization level 110 pts/atom (Figures S1 and S2).

The performance of the randomized Block-Jacobi Preconditioner is shown in Figure 8, using as an example a single point COSMO RHF/6-31G calculation on a model protein (PDB ID: 2KJM, 516 atoms). Because RBJ is a randomized algorithm, each data point stands for the averaged results of 50 runs with different random seeds (error bars corresponding to the variance are also shown). For this test case, RBJ with a block size of 100 reduces the total number of CG steps (matrix-vector products) by 40% compared to fixed threshold CG. Increasing the block size to 800 only slightly enhances the performance. As a reference, we also implemented the block Jacobi preconditioner based on the octree algorithm. In Figure 8, "octree-800" denotes the octree preconditioner with at most 800 points in each octree leaf box. Unlike RBJ, the number of points in each block of the octree is not fixed. For octree-800, the mean block size is 289. RBJ-100 already outperforms octree-800 in the number of CG steps, despite the smaller size of blocks, because RBJ provides better control of the block size and is less sensitive to the shape of the molecular surface. For RBJ and octree preconditioners with the same average blocksize $\bar{l}$, if the molecular shape is irregular (which is common for large asymmetric bio-molecules), the octree will contain both very small and large blocks for which $l \ll \bar{l}$ or $l \gg \bar{l}$, respectively. This effect reduces the efficiency of the octree algorithm in two ways: 1) the small blocks tend to be poor at preconditioning and 2) the large blocks are less efficiently stored and inverted.

Another important aspect of the preconditioner is the overhead. For a system with a small number of MS points (e.g. less than 1000), the time saved by reducing CG steps cannot compensate the overhead of building blocks for RBJ. Thus, a standard Jacobi preconditioner is faster. For a system with a large number of MS points, the RBJ preconditioner is significantly faster than Jacobi, despite some overhead for building and inverting the blocks. As shown in Figure 7, compared with the "fixed $\delta$ +Jacobi" method, "fixed $\delta$ +RBJ" provides a 1.5X speedup, and "dynamic $\delta$ + RBJ" provides a 3X speedup.

*4c. PCM gradient evaluation*

To efficiently evaluate Eq. (14), we note that $\nabla_{R_I}\mathbf{A}, \nabla_{R_I}\mathbf{B}$ and $\nabla_{R_I}\mathbf{c}$ are all sparse and do not need to be calculated explicitly for all nuclear coordinates. This is a direct result of the fact that each MS point only moves with the atom on which it is centered, which is also true for the basis functions.

Therefore, the strategy here is to only evaluate the non-zero terms and add them to the corresponding gradients. Specifically, we focus on the evaluation of the second term $\left(\nabla_{R_I}^*\mathbf{c}^\dagger\right)\mathbf{q}$ in Eq. (14), which involves one-electron integrals and is the most demanding.

For each interaction between an MS point and a primitive pair, there are three non-zero derivatives: $[\nabla_{R_I}\chi_i | \hat{J}_k^{screened} | \chi_j], [\chi_i | \hat{J}_k^{screened} | \nabla_{R_J}\chi_j], [\chi_i | \nabla_{R_K}\hat{J}_k^{screened} | \chi_j]$, where $\chi_i, \chi_j$ and MS point $k$ are located on atoms $I$, $J$ and $K$, respectively. Therefore, $(\nabla_{R_I}^*\mathbf{c}^\dagger)\mathbf{q}$ is composed of three parts

$$\begin{aligned}
(\nabla_{R_I}^*\mathbf{c}^\dagger)\mathbf{q} &= \sum_{ij, i\in I} ga[ij] + \sum_{ij, j\in I} gb[ij] + \sum_{k\in I} gc[k] \\
ga[ij] &= P_{\mu\nu}c_{\mu i}c_{\nu j}\sum_k q_k[\nabla_I\chi_i | \hat{J}_k^{screened} | \chi_j] \\
gb[ij] &= P_{\mu\nu}c_{\mu i}c_{\nu j}\sum_k q_k[\chi_i | \hat{J}_k^{screened} | \nabla_J\chi_j] \\
gc[k] &= q_k\sum_{ij} P_{\mu\nu}c_{\mu i}c_{\nu j}[\chi_i | \nabla_K\hat{J}_k^{screened} | \chi_j]
\end{aligned} \quad (22)$$

The calculation of *ga* and *gb* requires reduction over MS points, whereas *gc* requires reduction over primitive pairs. Therefore, the GPU algorithm for evaluation of $(\nabla^*_{R_I} \mathbf{c}^\dagger)\mathbf{q}$ is a hybrid of the pair-driven $\Delta \mathbf{F}^S$ kernel and the MS-driven **c** kernel. Primitive pairs are prescreened with the density-weighted Schwartz bound of Eq. (18). Each thread is assigned a single primitive pair, and loops over all MS points. Integrals *ga*[*ij*] and *gb*[*ij*] are accumulated within each thread. Finally, *gc*[*k*] is formed by a reduction sum within each block at the end of the *k*th loop, and the host CPU performs the cross-block reduction.

## 5. Performance

A primary concern is the efficiency of a COSMO implementation compared with its gas phase counterpart at the same level of *ab initio* theory. For our set of 20 proteins, Figure 9 shows the ratio of time consumed for COSMO compared to gas phase for RHF/6-31G single point energy calculations. The COSMO calculations introduce at most 60% overhead. A similar ratio is achieved for the calculation of analytic gradients (Figure S3). Of course, this ratio will change with the level of quantum chemistry method and MS discretization. For a medium-sized molecule, the ratio decreases as the basis set size increases (Figure S4) because the COSMO-specific evaluations only involve one-electron integrals, whose computational cost grows more slowly than that of the gas phase Fock build. The COSMO overhead also decreases as larger cavity radii are used (Figure S5), because the number of MS points decreases with increasing cavity radii (more points are buried in the surface). This trend is expected to apply to molecules in a wide range of sizes (ca. 80-1500 atoms), as they share a general trend of decreasing the number of MS points with increasing radii (Figure S6). As a specific example, we turn to the Photoactive Yellow Protein (PYP, 1537 atoms). When the most popular choice[55] of cavity radii (choosing atomic radii to be 20% larger than Bondi radii, i.e. 1.2*Bondi) is used (76577 MS points in total), the computational effort associated with COSMO takes approximately 25% of the total runtime for COSMO RHF/6-31G* single point calculation (Figure 10). When larger cavity radii (2.0*Bondi) are used (17266 MS points), the overhead for COSMO falls to 5% (Figure S7). Overall, our COSMO implementation typically requires about 20-40% more time than gas phase energy or gradient calculations, when a moderate basis set (6-31G) and typical cavity discretization level is used (radii=1.2*Bondi, 110 pts/atom). When a larger basis set or larger cavity radii is used, COSMO will cost less and be an even more insignificant part of the total computational cost relative to a gas phase calculation.

To demonstrate the advantage of a GPU-based implementation, we compare our performance to a commercially-available, CPU-based quantum chemistry code, Q-Chem.[34] We take the smallest (PDB ID: 1Y49, 122 atoms) and the largest (PDB ID: 2KJM, 516 atoms) molecules in our test set of proteins and run a RHF/6-31G COSMO-ISWIG gradient calculation. TeraChem calculations were run on nVidia GTX TITAN GPUs and Intel Xeon X5690@3.47 GHz CPUs. Q-Chem calculations were run on faster Intel Xeon ES-2643@3.30 GHz CPUs. The number of GPUs/CPUs was varied in the tests to assess parallelization efficiency across multiple CPU/GPUs.

Timing results are summarized in Table 1 and Table 2. The PCM gradient calculation consists of four major parts: gas phase SCF (SCF steps in common with gas phase calculations), PCM SCF (including building the **c** vector, building $\Delta \mathbf{F}^S$, and the CG linear solve), gas phase gradients, and PCM gradients. For each portion of the calculation, the runtime is annotated in parenthesis with the percentage of the runtime for that step relative to total runtime. As explained above, Q-Chem uses OpenMP with no matrix mode and FMM. Comparisons with the MPI parallelized version of Q-Chem are provided in the supporting information. The MPI version of Q-Chem does not use FMM and stores the A matrix explicitly.

First we focus on the single CPU/GPU performance, and we compare the absolute runtime values. For both the small and large systems, the GPU implementation provides a 16X reduction in the total runtime relative to Q-Chem. This is in spite of the fact that Q-Chem is using a linear scaling FMM method. The speedup for different sections varies. The PCM gradient calculation has a speedup of over 40X, which is much higher than the overall speedup and the speedup for gas phase gradient. The FMM-based CG procedure in Q-Chem is slower than the version which explicitly stores the **A** matrix. Even compared to the latter, our CG implementation is about 3X faster (see SI). We attribute this to the preconditioning and dynamic threshold strategies described above. On the other hand, it is interesting to note that Q-Chem and TeraChem both spend a similar percentage(22-27%) of their time on PCM SCF and gradient evalutions, regardless of the difference in absolute runtime.

When we use multiple GPUs/CPUs, the total runtime decreases as a result of parallelization for both Q-Chem and TeraChem. However, for both programs, the percentage of time spent on PCM increases, showing that the parallel efficiency of the PCM related evaluations is lower than that of other parts of the calculation. Table 3 shows the parallel efficiency of TeraChem PCM calculation. The parallel efficiency is defined here as usual:[56]

$$\text{efficiency} = \frac{1}{P}\frac{T_1}{T_P} \qquad (23)$$

where P is the number of GPUs/CPUs in use and $T_1/T_P$ are the total runtime in serial/parallel, respectively. We compare the parallel efficiency of the four components of the PCM SCF calculation: building **c**, building $\Delta \mathbf{F}^S$, solving CG, and building the other terms in common with gas phase SCF. The parallel efficiencies of building **c** and $\Delta \mathbf{F}^S$ are both higher than that of gas phase SCF. However, for our CG implementation, the matrix-vector product is calculated on the CPU, which hampers the overall PCM SCF parallel efficiency. Similarly, parallel efficiency of the PCM gradient evaluation is limited by our serial computation of $\nabla \mathbf{A}, \nabla \mathbf{B}$.

Overall, the GPU implementation of PCM calculations in TeraChem demonstrates significant speedups compared to Q-Chem, which serves as an example of the type of performance expected from a mature and efficient CPU-based COSMO implementation. However, our current implementations of CG and $\nabla \mathbf{A}, \nabla \mathbf{B}$ are conducted in serial on the CPU and do not benefit from parallelization. This is a direction for future improvement.

## 6. Applications

As a representative application, we studied the structure of a protein fibril[57] (protein sequence SSTVNG, PDB ID: 3FTR) with our COSMO code. This fibril is known to be able to form dimers called "steric zippers" that can pack and form amyloids -- insoluble fibrous protein aggregates. In each "zipper" pair, the two segments are tightly interdigitated β-sheets with no water molecules in the interface. The experimental structure of SSTVNG is a piece of the zipper from a fibril crystal. Kulik et al.[39] found that minimal basis set *ab initio*, gas phase, geometry op-timizations of a zwitterionic 3FTR monomer resulted in a structure with an unusual deprotonation of amide nitrogen atoms. In that structure, the majority of the amide protons are shared between peptide bond nitrogen atoms and oxygen atoms, forming a covalent bond with the oxygen and a weaker hydrogen bond with the nitrogen. This phenomenon was explained as an artifact caused by both the absence of surrounding solvent and the minimal basis set. We were interested to quantify the degree to which these two approximations affected the outcome. Thus, we conducted more expansive geometry optimizations of 3FTR with and without COSMO to investigate how solvation influences the conformational landscape of the protein.

Stationary point structures of 3FTR were obtained as follows: starting from the two featured structures found previously (an unusually protonated structure and a normally protonated stationary point structure close to experiment), geometry optimizations were conducted in gas phase and with COSMO to describe aqueous solvation ($\varepsilon=78.39$). Whenever a qualitatively different structure was encountered, that structure was set as a new starting point for geometry optimization under all levels of theory. Through this procedure, seven different types of stationary point structures were found (Figures 11 and 12 and Table S3) characterized by differing protonation states and backbone structures. We characterize the backbone structure by the end-to-end distance of the protein, computed as the distance between the $C_\alpha$ atoms of the first and last residue. We describe the protonation state of the amide N and O with a "protonation score," defined as follows:

$$\text{Protonation Score} = \frac{\sum_{i=1}^{n_r} d_{O_i-H_i} / d_{N_i-H_i}}{n_r} \qquad (24)$$

where $n_r$ is the number of residues; $O_i$, $H_i$, and $N_i$ represent the amide O, H, N belonging to the $i$th residue (for the 1st residue, $H_i$ represents the hydrogen atom at the N-terminus of the peptide closest to O). The higher the score is (e.g. > 1.5), the more closely hydrogens are bonded with amide nitrogens, indicating a correct protonation state.

The 3FTR crystal structure is zwitterionic with charged groups at both ends, and geometry optimized structures of isolated 3FTR peptides will find minima that stabilize those charges. In the gas phase, the zwitterionic state's energy is lowered during geometry minimizations in two ways. In one case, the C-terminus carboxylate is neutralized by a proximal amide H, resulting in unusually protonated local minima. In the other case, the energy is minimized by backbone folding which brings the charged ends close to each other. Both rearrangements result in unexpected structures inconsistent with experiments in solution. We note however that such structural rearrangements are known to occur in gas phase polypeptides.[58]

COSMO solvation largely corrects the protonation artifact observed in gas phase. Two types of severely unusually protonated (protonation score <1.5) local minima are observed. One (labeled min1u in Figures 11 and 12) has been previously reported with the straight backbone structure as crystal structure. The other unusually protonated local minimum is min2u, which has very similar protonation state as min1u, but a slightly bent backbone (backbone length <17Å). The normally protonated counterparts of min1u and min2u are min1n and min2n, which are the two minima most resembling the crystal structure. In gas phase calculations with 3-21G and 6-31G, these four minima are all over 50 kcal/mol higher in energy than a folded structure (min4). COSMO solvation stabilizes min1n and min2n by about 50 kcal/mol, while leaving the anomalous min1u and min2u as high-energy structures (Table 4, Figure 11 and Figure 12). Moreover, this COSMO stabilization effect is already quite large for the smallest basis set (COSMO stabilization for different basis sets is summarized in Table 4). Although min1u and min2u are still preferred over the normally protonated structures in both gas phase and COSMO STO-3G calculations, this is perhaps expected since the basis set is so small.

COSMO also plays an important role in stabilizing an extended backbone structure. In gas phase calculations, the larger the end-to-end distance is, the less stable the structure tends to be. For both RHF/6-31G and ωPBEh calculations (Figure 11 and Figure 12, respectively), all unfolded structures (min1n, min1u, min2n, min2u, min2t) are very unstable in the gas phase with respect to the folded structure, min4. Among them, min1n and min2n have the largest charges separated by the largest distances (Table S6). COSMO stabilizes the terminal charges, thus significantly lowering the energy of min1 and min2. For COSMO RHF/6-31G, min2n is as stable as the folded min4. At the same time, the half folded and twisted structure, min3, is destabilized by COSMO.

For the most part, the local minima in the gas phase and solution are similar for this polypeptide, even across a range of basis sets including minimal sets. However, the relative energies of these minima are strongly affected by solvation and basis set. Solvation is especially important in this case because of the zwitterionic character of the polypeptide. This is expected on physical grounds (and the structures of gas phase polypeptides and proteins likely reflect this), and strongly suggests that solvation effects need to be modeled when using *ab initio* methods to describe protein structures.

## 7. Conclusions

We have demonstrated that by implementing COSMO-related electronic integrals on GPUs, dynamically adjusting the CG threshold for COSMO equations, and applying a new strategy for generating the block Jacobi preconditioner, we can significantly decrease the computational effort required for COSMO calculations of large biomolecular systems. We achieve speedups compared to CPU-based codes of more than 15-60X. The computational overhead introduced by the COSMO calculation (relative to gas phase calculations) is quite small – typically 20-40%. Finally, we showed an example where COSMO solvation influences the geometry optimization of proteins qualitatively. Our efficient implementation of COSMO will be useful for the study of protein structures.

Our approach for COSMO electron integral evaluation on GPU can be adapted for other variants of PCMs, such as the integral equation formalism (IEF-PCM or SS(V)PE).[59] Since generation of the randomized block Jacobi preconditioner only depends on the matrix itself (not the specific physical model used), the strategy can be applied to the preconditioning of CG in a variety of fields. For instance, for linear scaling SCF, an alternative to diagonalization is the direct minimization of the energy functional[60] with preconditioned CG. Another example is the solution of a large linear system with CG to obtain the perturbative correction to the wavefunction in CASPT2.[61]

In the future we will extend our acceleration strategies to non-equilibrium solvation, where the optical (electronic) dielectric constant is equilibrated with the solute while the orientational dielectric constant is not.[62-64] This will allow modeling of biomolecules in solution during photon absorption, fluorescence and phosphorescence processes. Our accelerated PCM code will also facilitate calculation of redox potential of metal complexes[65] in solutes and pKa values for large biomolecules.[66]

## 8. Acknowledgments

This work was supported by the AMOS program within the Chemical Sciences, Geosciences, and Biosciences Division of the Office of Basic Energy Sciences, Office of Science, US Department of Energy. TJM is grateful to the Department of Defense (Office of the Assistant Secretary of Defense for Research and Engineering) for a National Security Science and Engineering Faculty Fellowship (NSSEFF).

**Table 1.** Timing data (seconds) for COSMO RHF/6-31G gradient calculation of TeraChem (TC) on GTX TITAN GPUs and Q-Chem (QC) on Intel Xeon CPUs ES-2643 @ 3.30 GHz.

| molecule (#atoms, #MS points) | #GPU /CPU core | Total runtime | | | PCM gradient | | | Gas phase gradient | | | PCM SCF | | | Gas Phase SCF | | |
|---|---|---|---|---|---|---|---|---|---|---|---|---|---|---|---|---|
| | | QC | TC | speed-up | QC | TC | speed-up | QC | TC | speed-up | QC | TC | speed-up | QC | TC | speed-up |
| 1y49 (122,5922) | 1 | 1877 | 116 | 16.2 | 88 (5%) | 2 (2%) | 39.4 | 410 (22%) | 22 (19%) | 18.5 | 502 (27%) | 25 (22%) | 19.9 | 878 (47%) | 66 (57%) | 13.3 |
| | 4 | 705 | 40 | 17.5 | 85 (12%) | 2 (4%) | 56.8 | 84 (12%) | 6 (15%) | 14.1 | 337 (48%) | 10 (25%) | 33.5 | 200 (28%) | 23 (57%) | 8.7 |
| | 8 | 581 | 31 | 18.9 | 89 (15%) | 1 (4%) | 67.2 | 72 (12%) | 4 (12%) | 20.2 | 309 (53%) | 8 (27%) | 36.8 | 111 (19%) | 17 (57%) | 6.3 |
| 2kjm (516,26025) | 1 | 35345 | 1787 | 19.8 | 1960 (6%) | 40 (2%) | 48.9 | 6840 (19%) | 417 (23%) | 16.4 | 7789 (22%) | 445 (25%) | 17.5 | 18756 (53%) | 885 (50%) | 21.2 |
| | 4 | 13506 | 623 | 21.7 | 2100 (16%) | 26 (4%) | 79.6 | 1415 (10%) | 116 (19%) | 12.2 | 6043 (45%) | 181 (29%) | 33.4 | 3948 (29%) | 299 (48%) | 13.2 |
| | 8 | 11339 | 419 | 27.0 | 2088 (18%) | 23 (6%) | 89.2 | 1144 (10%) | 59 (14%) | 19.4 | 5768 (51%) | 141 (33%) | 41.1 | 2339 (21%) | 196 (47%) | 11.9 |

**Table 2.** As in Table 1, but detailed information for timing in the COSMO portion of the calculation.

| molecule (#atoms, #MS points) | #GPU/CPU | CG | | | Build **c** | | | Build $\Delta \mathbf{F}^s$ | | |
|---|---|---|---|---|---|---|---|---|---|---|
| | | QC | TC | speed-up | QC | TC | speed-up | QC | TC | speed-up |
| 1y49 (122, 5922) | 1 | 221.3 (12%) | 4.4 (4%) | 39.8 | 149.5 (8%) | 8.9 (8%) | 16.8 | 131.0 (7%) | 10.1 (9%) | 13.0 |
| | 4 | 55.5 (8%) | 3.7 (9%) | 15.4 | 149.6 (21%) | 2.6 (7%) | 56.7 | 131.5 (19%) | 3.1 (8%) | 42.4 |
| | 8 | 28.0 (5%) | 3.8 (12%) | 7.5 | 149.6 (26%) | 2.1 (7%) | 70.9 | 131.6 (23%) | 1.9 (6%) | 70.6 |
| 2kjm (516, 26025) | 1 | 2335.1 (7%) | 97.6 (5%) | 18.8 | 2914.2 (8%) | 130.9 (7%) | 22.3 | 2539.3 (7%) | 175.8 (10%) | 14.4 |
| | 4 | 581.9 (4%) | 78.7 (13%) | 7.2 | 2918.9 (22%) | 38.5 (6%) | 75.8 | 2541.7 (19%) | 48.1 (8%) | 52.8 |
| | 8 | 310.8 (3%) | 78.0 (19%) | 3.8 | 2918.2 (26%) | 20.2 (5%) | 144.5 | 2538.7 (22%) | 25.4 (6%) | 100.1 |

**Table 3.** Parallel efficiency of TeraChem PCM calculation

|  | #GPU | PCM SCF | | | | Gas Phase SCF | PCM Gradient | | Gas phase Gradient |
|---|---|---|---|---|---|---|---|---|---|
|  |  | CG | Build c | Build $\Delta F^s$ | Total |  | $\nabla c$ | Total |  |
| 1y49 | 4 | 0.39 | 0.84 | 0.81 | 0.66 | 0.72 | 0.75 | 0.37 | 0.93 |
|  | 8 | 0.19 | 0.53 | 0.68 | 0.40 | 0.47 | 0.61 | 0.21 | 0.78 |
| 2kjm | 4 | 0.39 | 0.85 | 0.91 | 0.64 | 0.74 | 0.85 | 0.38 | 0.90 |
|  | 8 | 0.19 | 0.81 | 0.87 | 0.43 | 0.56 | 0.79 | 0.21 | 0.88 |

Table 4. Energy difference (kcal/mol) between the normally and unusually protonated 3FTR minima.

| Method/Basis Set | Energy Difference (kcal/mol) | | | |
|---|---|---|---|---|
| | $\Delta E(min1u-min1n)^a$ | | $\Delta E(min2u-min2n)^b$ | |
| | COSMO | Gas Phase | COSMO | Gas Phase |
| RHF-D/-STO-3G | -101 | -178 | -31 | -77 |
| RHF/STO-3G | -106 | -179 | -27 | -76 |
| RHF/3-21G | 77 | 13 | 83 | 6 |
| RHF/6-31G | 90 | 29 | 102 | 13 |

[a] min1n and min1u are minima with an extended backbone structure (as in the 3FTR crystal structure), where 'n' stands for normal protonation state, and 'u' stands for 'unusual' protonation state.
[b] min2n and min2u are minima with slightly bent backbone structure.

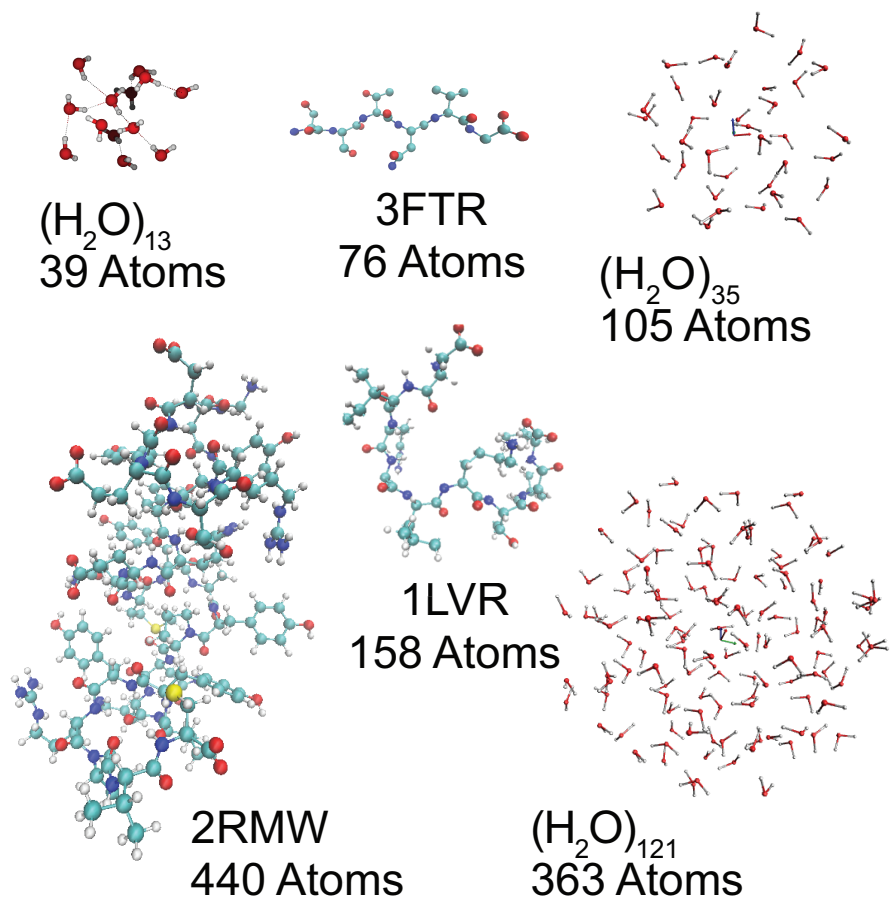

**Figure 1**. Molecular geometries used to benchmark the correlation between COSMO energy error and CG convergence threshold.

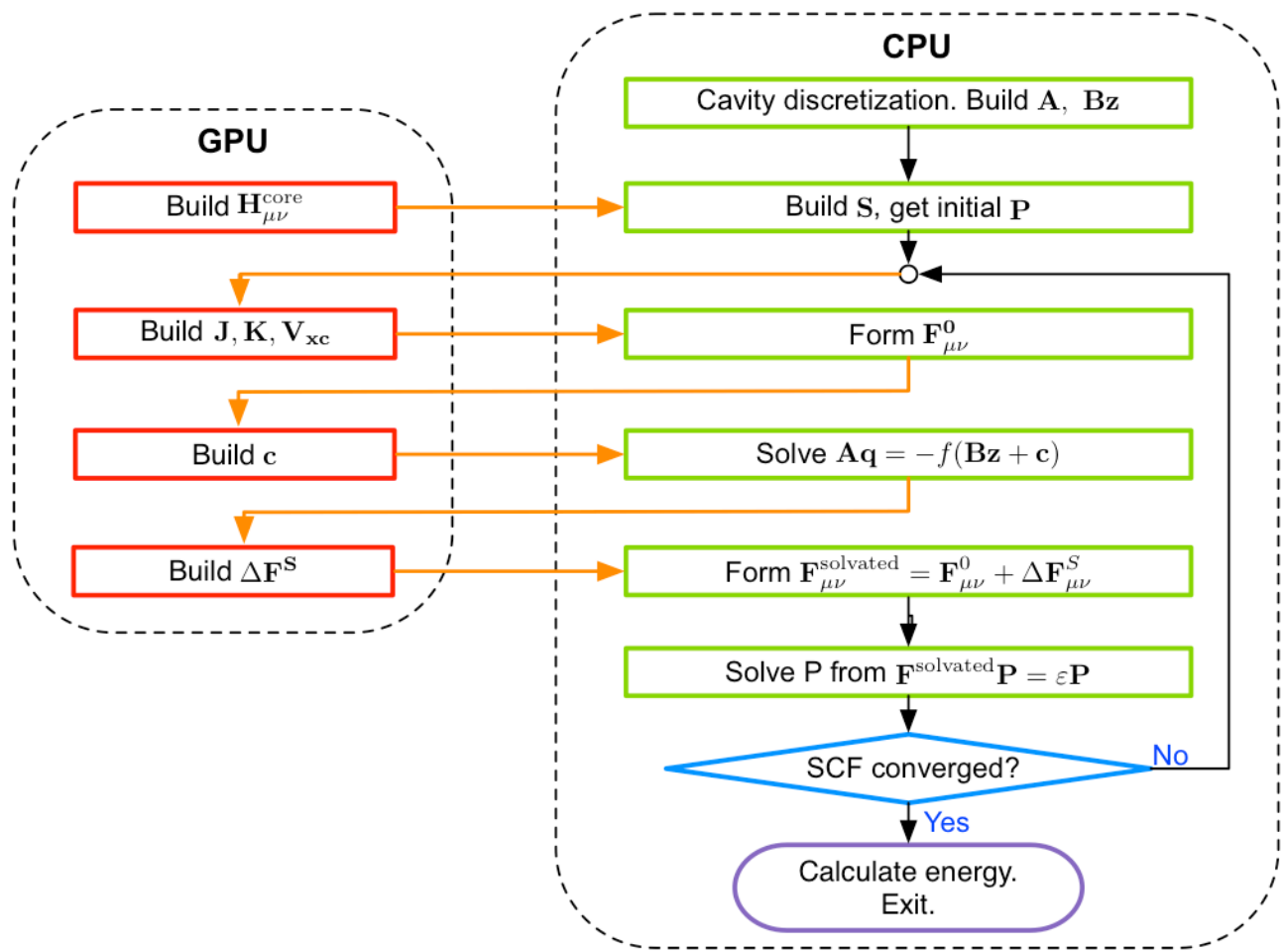

**Figure 2**. Flowchart for COSMO SCF implementation

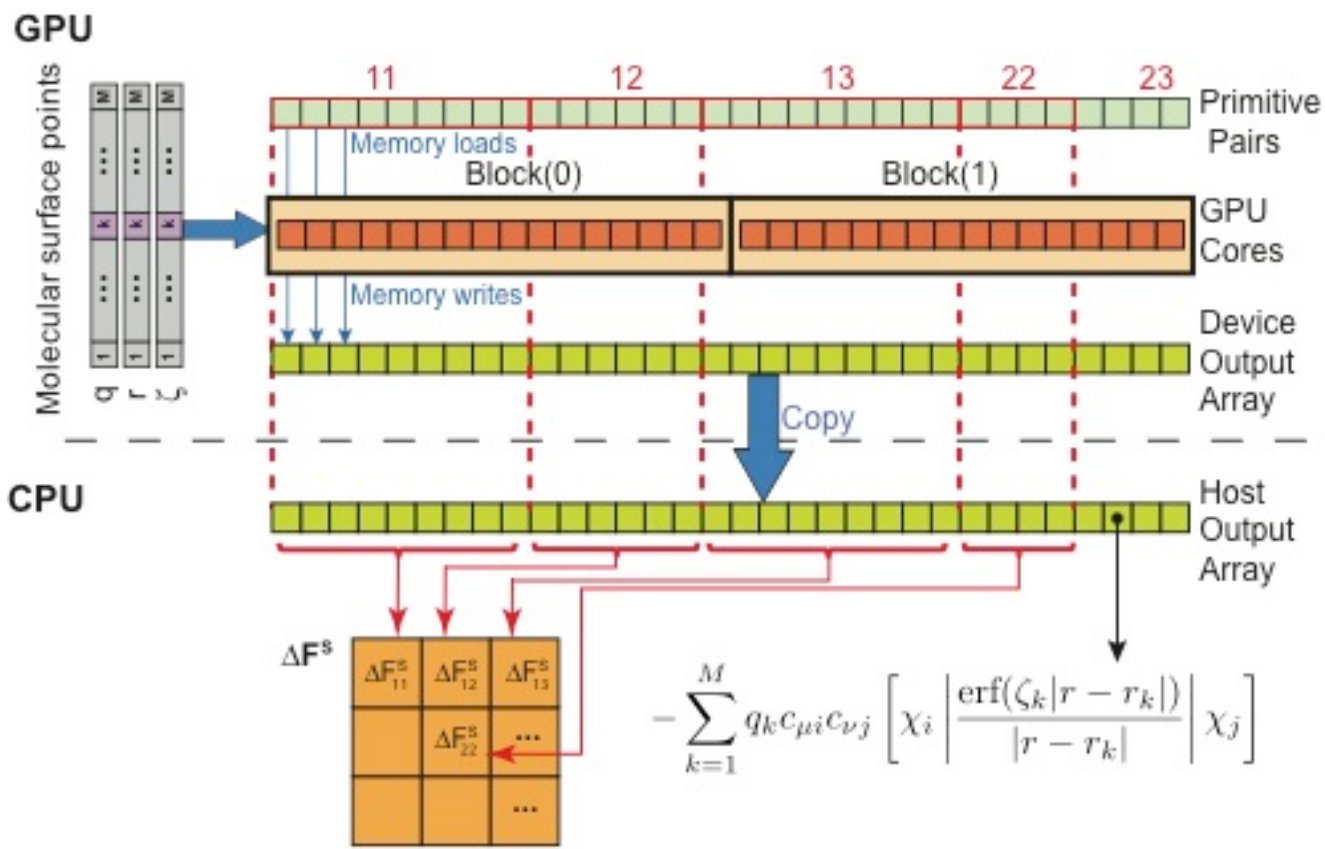

**Figure 3**. Algorithm for calculating $\Delta F^S$ for *ss* integrals of a system compsed of 3 *s* shells (the first and the third *s* shells contain 3 primitive Gaussian function each. The second *s* shell has 2 primitive Gaussian functions). On top of the graph, the pale green array represents primitive pairs belonging to *ss* shell pairs. The GPU cores are represented by orange squares (threads) embedded in pale yellow rectangles (1 dimensional blocks with 16 threads/block). The output is an array where each entry stores a primitive pair integral. Primitive pair integrals are finally added to the Fock matrix entry of the conrresponding contracted function pair. All red lines and text indicate contracted Gaussian integrals. Blue arrows and text indicate memory operations.

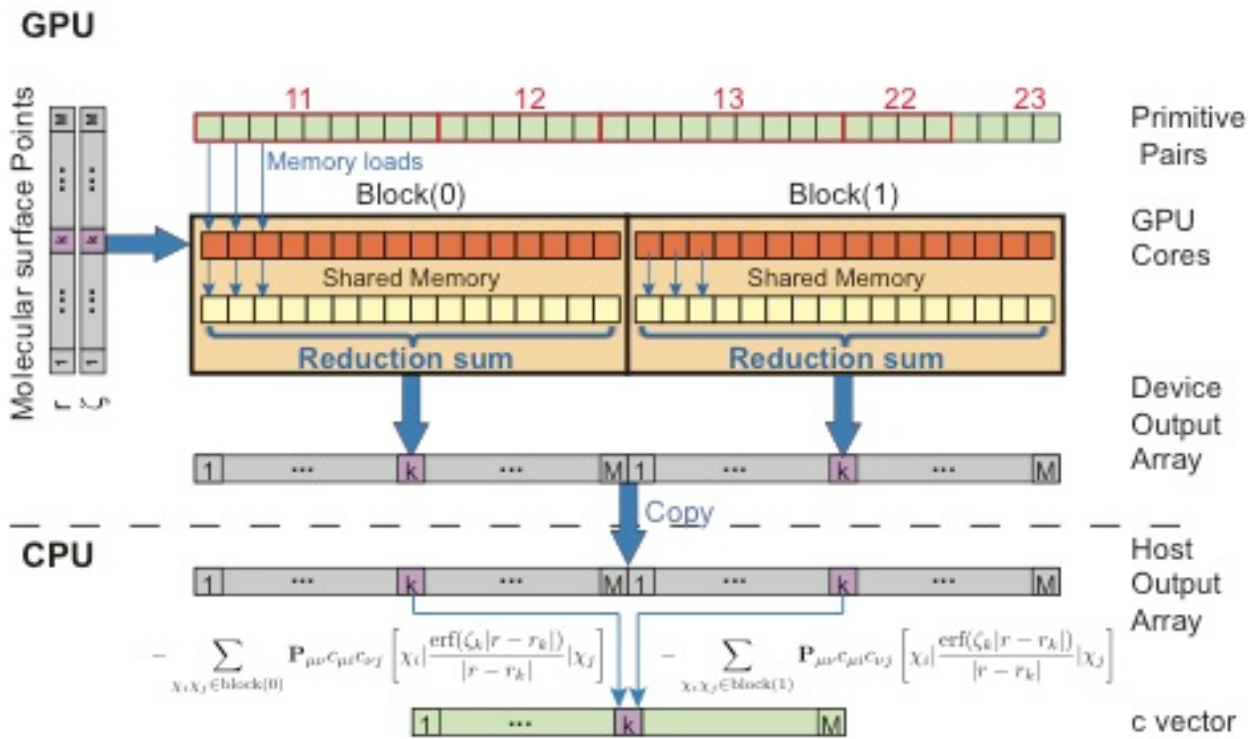

**Figure 4**. MS point-driven algorithm for building **c** for *ss* integrals of a system composed of 3 *s* shells (the first and the third *s* shells contain 3 primitive Gaussian functions each. The second *s* shell has 2 primitive Gaussian functions). The pale green array at the top of the figure represents primitive pairs belonging to *ss* shell pairs. The GPU cores are represented by orange squares (threads) embedded in pale yellow rectangles (1 dimensional blocks with 16 threads/block). The output is an array where each entry stores a primitive pair integral. Primitive pair integrals are finally added to the Fock matrix entry of the conrresponding contracted function pair. All red lines and text indicate contracted Gaussian integrals. Blue arrows and text indicate memory operations.

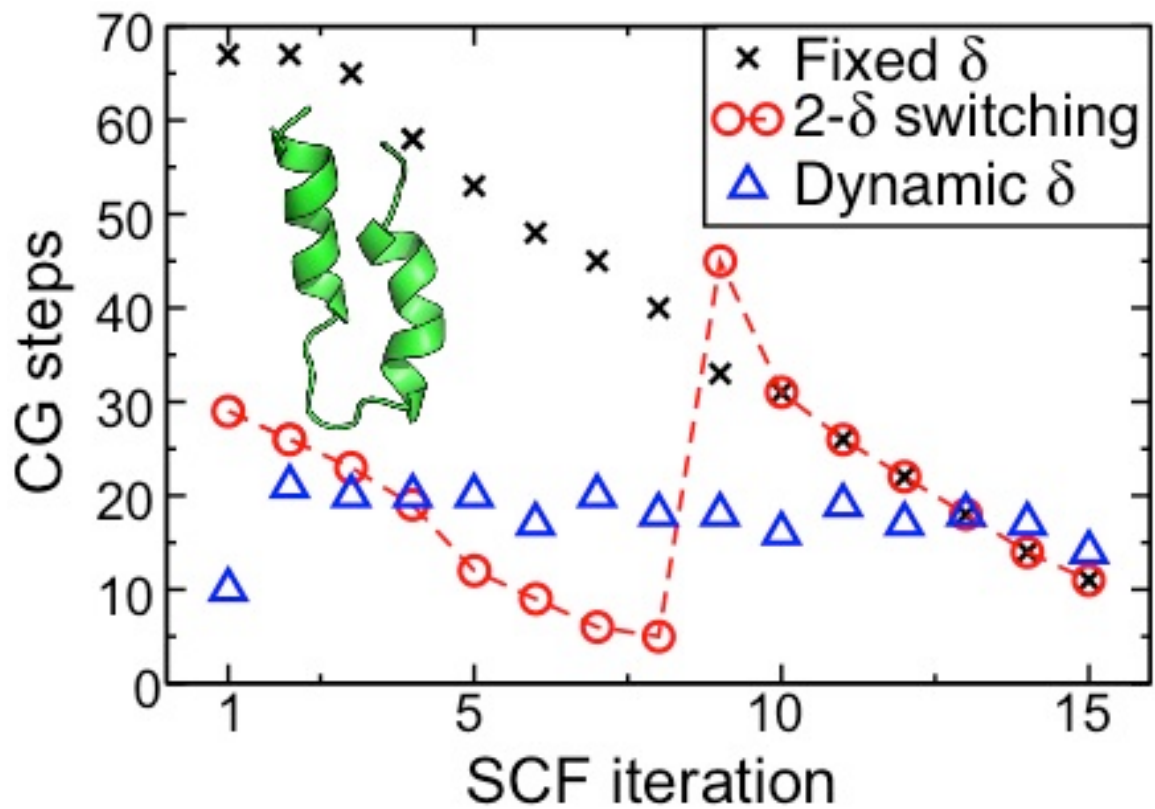

**Figure 5**. Number of CG steps taken in each SCF iteration for different CG residual convergence threshold schemes in COSMO RHF/6-31G calculation on a model protein (PDB ID: 2KJM, 516 atoms, shown in inset).

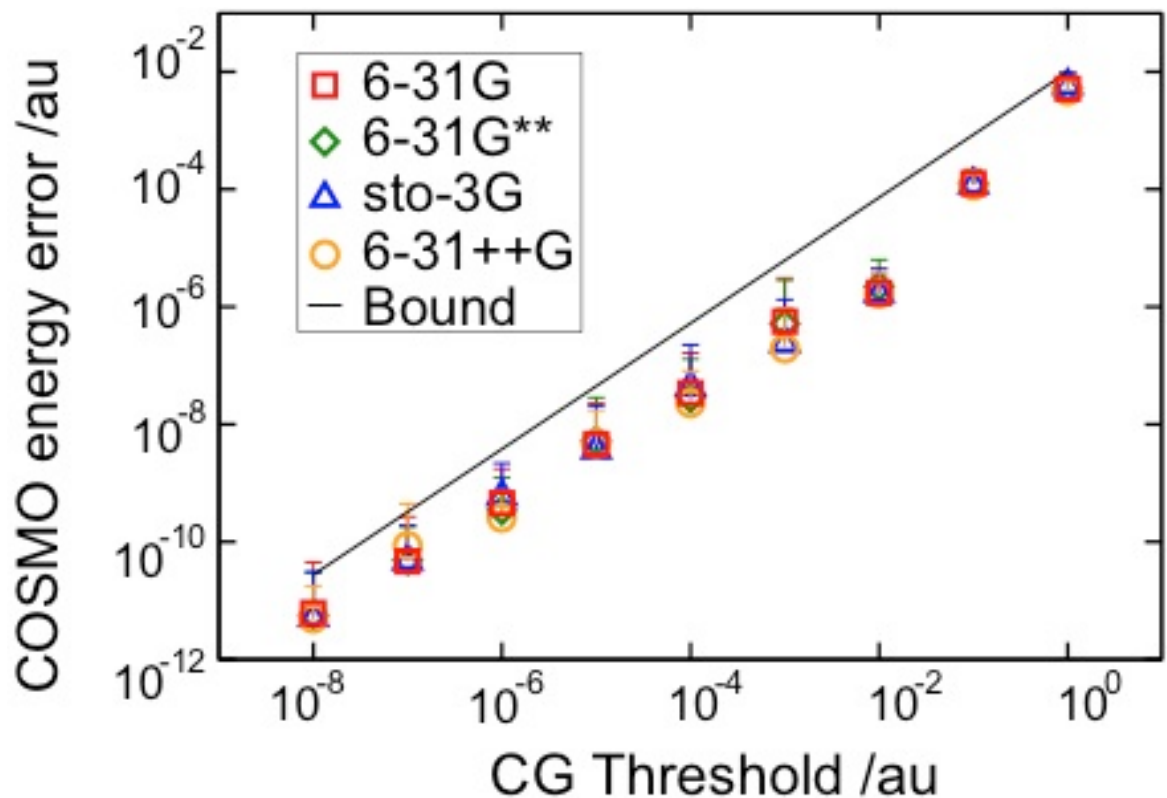

**Figure 6**. Average absolute error in first COSMO energies versus the CG residual convergence threshold. Both minimized and distored nonequilibrium geometries for the the test set are included in averages. Error bars represent two standard deviations above the mean. The black line represents the empirical error bound given by Eq. (21).

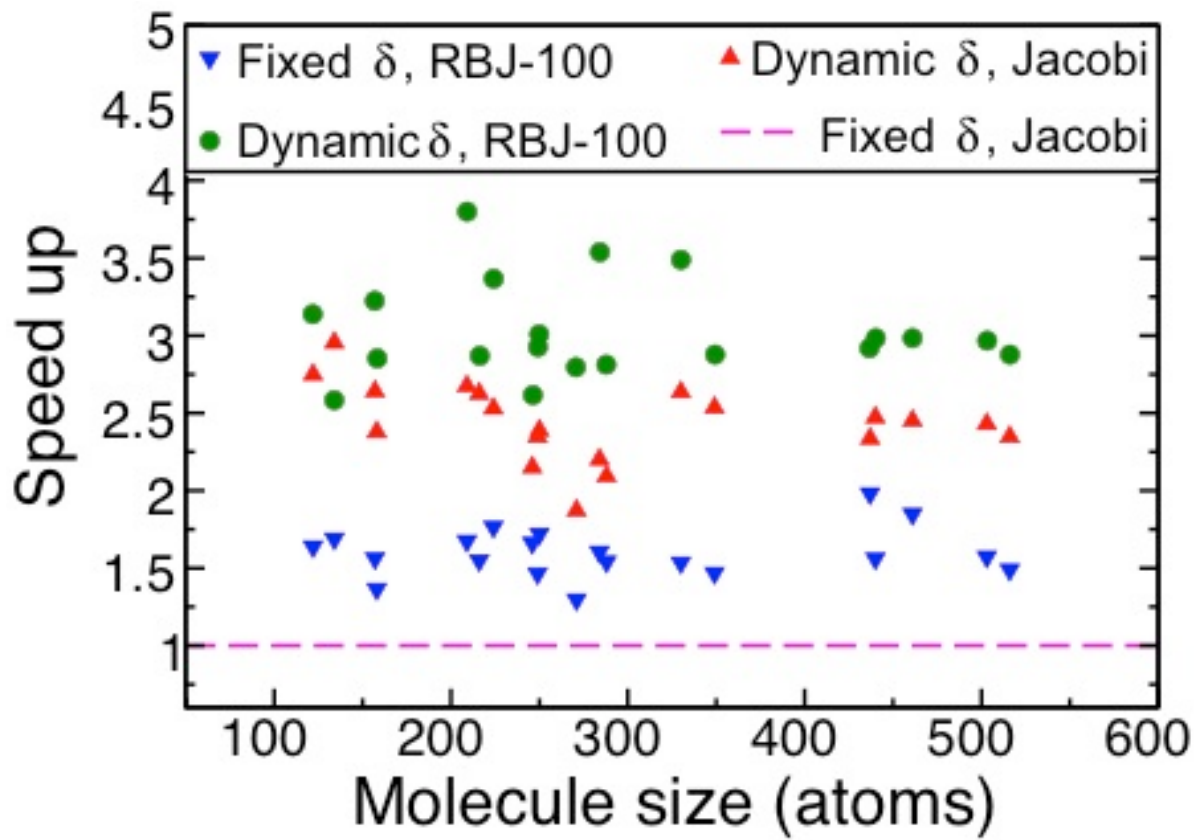

**Figure 7**. Speed up for CG linear solve methods compared to fixed δ + Jacobi preconditioner of TeraChem for COSMO RHF/6-31G single point energy calculations. Calculations were carried out on 1 GPU (GeForce GTX TITAN).

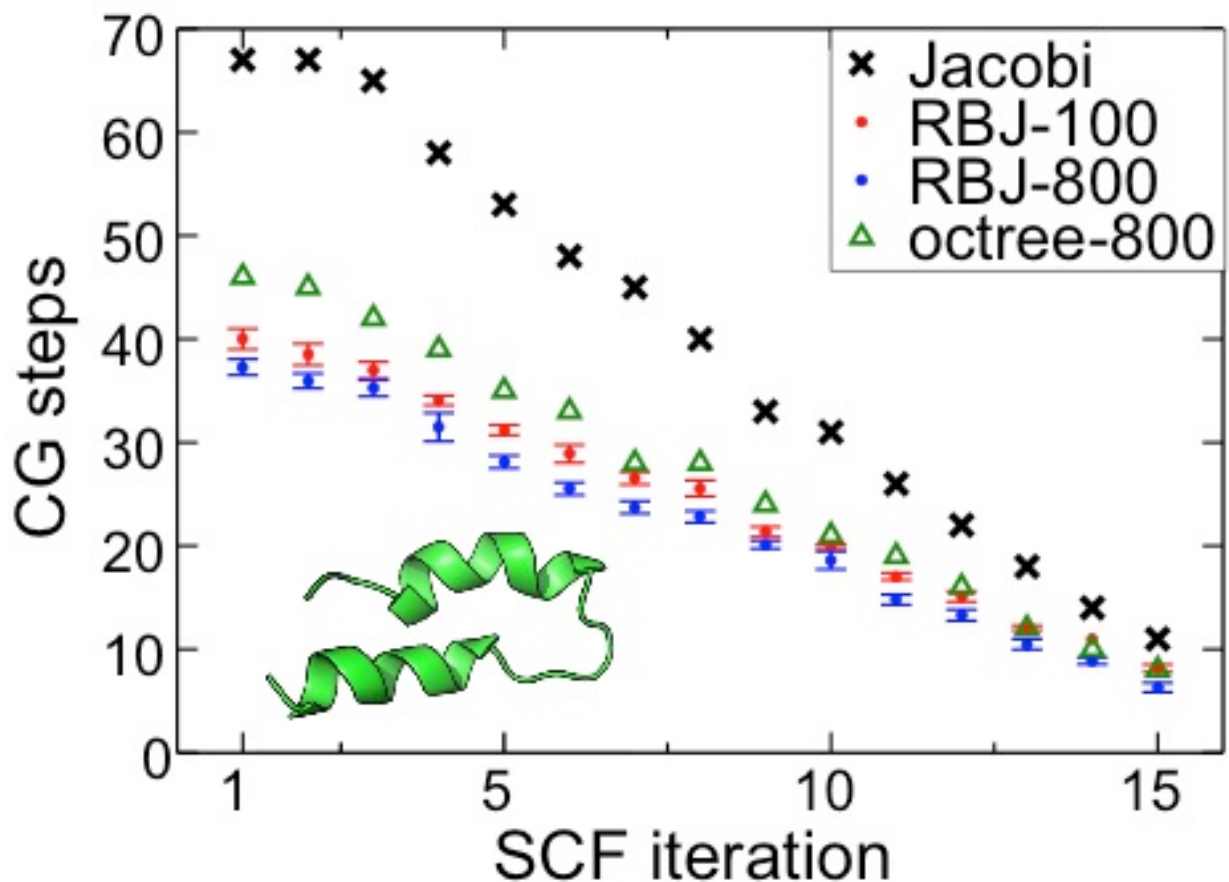

**Figure 8**. Number of CG steps taken in each SCF iteration for different choices of CG preconditioner in COSMO RHF/6-31G calculation on a model protein (PDB ID: 2KJM, 516 atoms, shown in inset). RBJ-100 and RBJ-800 represent the randomized block Jacobi preconditioner with block size of 100 and 800, respectively. The block Jacobi preconditioner based on an octree partition of surface points (denoted octree-800) is also shown, where the maximum number of points in a box is 800.

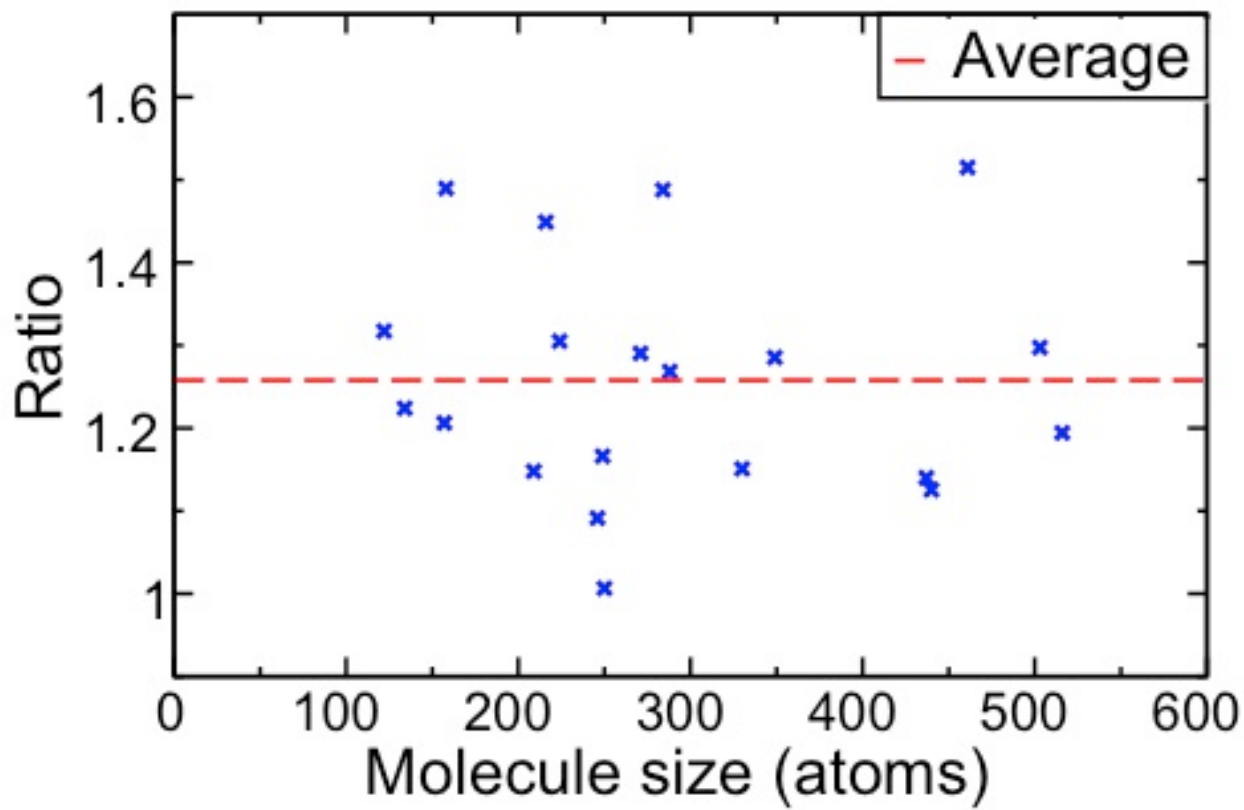

**Figure 9.** Ratio of time for COSMO versus gas phase single point energy calculation for 20 small proteins at RHF/6-31G level. Dynamic precision for 2-electron integrals is used with COSMO cavity radii chosen as 1.2*Bondi radii. An ISWIG discretization scheme is used with 110 Lebedev points/atom.

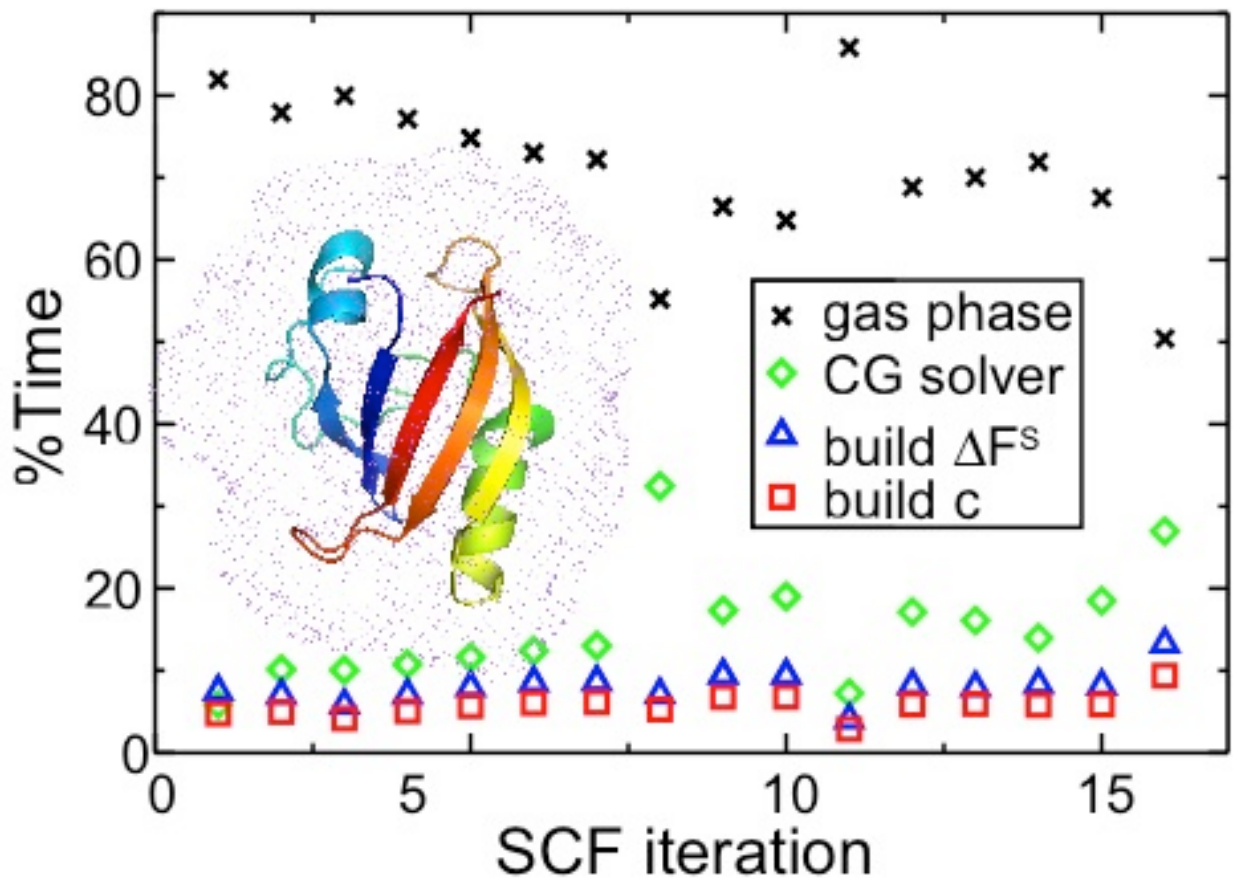

**Figure 10**. Breakdown of timings by SCF iteration for components of COSMO RHF/6-31G* calculation on Photoactive Yellow Protein (PYP) with cavity radii chosen as Bondi radii scaled by 1.2.

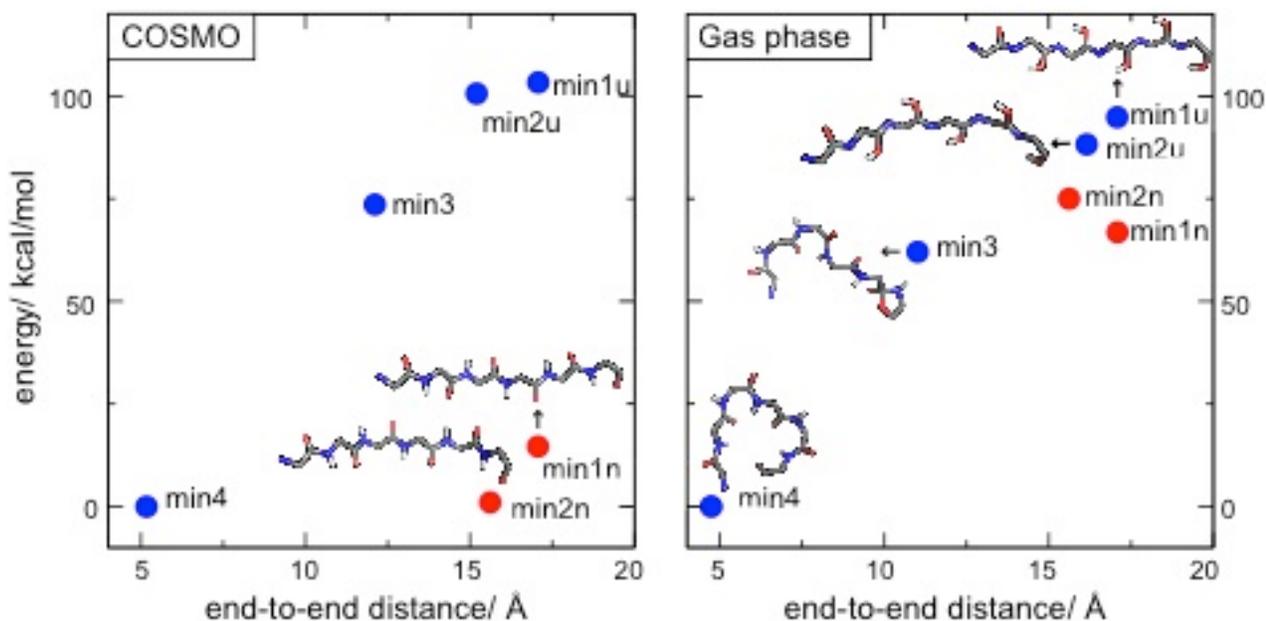

**Figure 11.** Different minima (min1n, min1u, min2n, min2u, min3, min4) of 3FTR found with RHF/6-31G geometry optimizations in COSMO and in the gas phase. The x-axis is the collective variable that characterizes the backbone folding. The y-axis is the total energy including solvation energy of the geometries. Each optimized structure is represented by a symbol in the graph and labeled by name with the backbone structure (C, O, N and H are colored grey, red, blue and white). Sidechains are omitted for clarity.

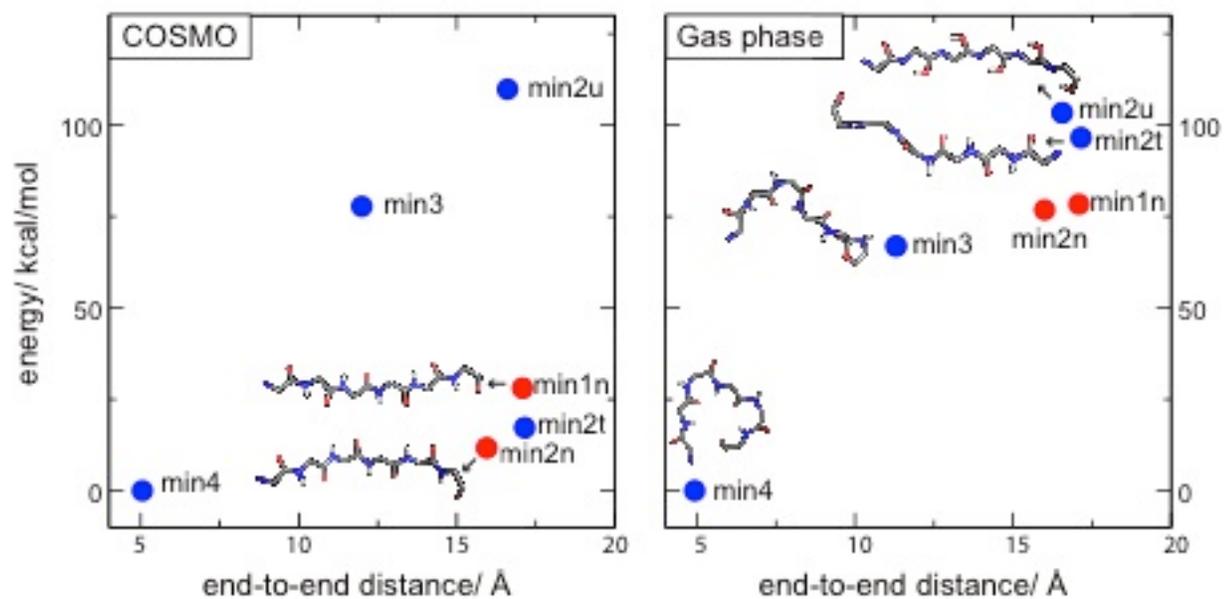

**Figure 12**. As in Figure 11, but using ωPBEh/6-31G.